\documentclass[preprint]{aastex}
\usepackage{amstext}
\usepackage{amssymb}
\usepackage{graphicx}
\usepackage{url}

\shorttitle{X-ray Observations of High-B Radio Pulsars}
\shortauthors{Olausen et al.}

\begin{document}

\title{X-ray Observations of High-B Radio Pulsars}

\author{S. A. Olausen\altaffilmark{1}, W. W. Zhu\altaffilmark{2}, J. K.
Vogel\altaffilmark{3}, V. M. Kaspi\altaffilmark{1}, A. G. Lyne\altaffilmark{4},
C. M. Espinoza\altaffilmark{4}, B. W. Stappers\altaffilmark{4},
R. N. Manchester\altaffilmark{5}, \& M. A. McLaughlin\altaffilmark{6}}

\altaffiltext{1}{Department of Physics, Rutherford Physics Building, McGill University,
3600 University Street, Montreal, Quebec, H3A 2T8, Canada}

\altaffiltext{2}{Department of Physics and Astronomy, University of British Columbia,
6224 Agricultural Road, Vancouver, BC, V6T 1Z1, Canada}

\altaffiltext{3}{Lawrence Livermore National Laboratory, 7000 East Avenue, Livermore,
CA 94550, USA}

\altaffiltext{4}{Jodrell Bank Centre for Astrophysics, School of Physics and Astronomy,
University of Manchester, Manchester, M13 9PL, UK}

\altaffiltext{5}{Australia Telescope National Facility, CSIRO Astronomy and Space
Science, Epping, NSW 1710, Australia}

\altaffiltext{6}{Department of Physics, West Virginia University, White Hall, Morgantown, WV 26506, USA}

\begin{abstract}
The study of high-magnetic-field pulsars is important for examining
the relationships between radio pulsars, magnetars, and X-ray-isolated
neutron stars (XINSs). Here we report on X-ray observations of three
such high-magnetic-field radio pulsars. We first present the results
of a deep \emph{XMM-Newton} observation of PSR J1734$-$3333, taken
to follow up on its initial detection in 2009. The pulsar's spectrum
is well fit by a blackbody with a temperature of $300\pm60$\,eV,
with bolometric luminosity $L_{\mathrm{bb}}=2.0_{-0.7}^{+2.2}\times10^{32}\,\mathrm{erg\, s^{-1}}\approx0.0036\dot{E}$
for a distance of 6.1\,kpc. We detect no X-ray pulsations from the
source, setting a $1\sigma$ upper limit on the pulsed fraction of
60\% in the 0.5--3\,keV band. We compare PSR J1734$-$3333 to other
rotation-powered pulsars of similar age and find that it is significantly
hotter, supporting the hypothesis that the magnetic field affects
the observed thermal properties of pulsars. We also report on \emph{XMM-Newton}
and \emph{Chandra} observations of PSRs B1845$-$19 and J1001$-$5939.
We do not detect either pulsar, setting $3\sigma$ upper limits on
their blackbody temperatures of 48 and 56~eV, respectively. Despite
the similarities in rotational properties, these sources are significantly
cooler than all but one of the XINSs, which we attribute to the two
groups having been born with different magnetic fields and hence evolving
differently.
\end{abstract}


\section{Introduction}

Over the past few decades, the boom in X-ray and gamma-ray astronomy
has led to a significant increase in our knowledge about the neutron
star family. Previously, the only kind of isolated neutron stars known
were radio pulsars. The latter are also known as `rotation-powered
pulsars' (RPPs) because their luminosities are generally much lower
than $\dot{E}$, the rate of energy loss due to rotation. Since then,
X-ray observations have led to the discovery of several new classes
of isolated neutron stars, including magnetars and X-ray-isolated
neutron stars (XINSs%
\footnote{Also known as X-ray dim isolated neutron stars (XDINSs).%
}); see \citet{k10} for a review. All of these classes exhibit properties
different from those of conventional RPPs.

Magnetars, including the soft gamma repeaters (SGRs) and anomalous
X-ray pulsars (AXPs), are isolated, slowly rotating neutron stars.
The known objects%
\footnote{See the McGill SGR/AXP Online Catalog at \url{http://www.physics.mcgill.ca/~pulsar/magnetar/main.html}.%
} have spin periods in the range of 2--12\,s, X-ray luminosities that
are in many cases much higher than their spin-down luminosities, and,
assuming conventional magnetic dipole braking, typically very high
magnetic fields $\left(B\sim10^{14}\textrm{--}10^{15}\,\mathrm{G}\right)$.
It is generally believed that their X-ray luminosities are fueled
by the decay of these high magnetic fields \citep{td95,td96,tlk02}.
Magnetars also show great X-ray variability and activity, ranging
from short SGR-like X-ray/$\gamma$-ray bursts to major, slow-rise
and long-lived X-ray flares, that are thought to be produced by fracturing
of the crust and reconfiguration of the magnetic field lines. For
reviews of magnetars, see for example \citet{wt06} or \citet{m08}.

The XINSs (see \citealt{h07} and \citealt{t09} for reviews) are
a group of nearby (distance $\leq500$\,pc) neutron stars characterized
by soft thermal X-ray spectra and no detected radio emission. Those
XINSs with detailed timing measurements have spin periods in the range
of 3--11\,s, high inferred magnetic fields $\left(B\sim\left(1\textrm{--}3\right)\times10^{13}\,\mathrm{G}\right)$,
and characteristic ages of $\sim$$10^{6}$\,yr \citep{kv09,kv11,vk08}.
The XINSs are considerably hotter and more X-ray luminous than similarly
aged ordinary RPPs \citep{kv09}, although they show no bursting activity
and are less luminous than the magnetars. Therefore, it has been proposed
\citep{kv09,pmg09,ppm+10} that the cooling of XINSs is affected by
heating of the neutron-star surface due to magnetic-field decay. If
correct, this suggests that XINSs may be evolutionarily related to
the magnetars; one possibility is that XINSs are magnetar descendants,
being older versions of neutron stars that were born with $10^{14}$--$10^{15}$\,G
magnetic fields which have slowly decayed with time.

The hint of magnetic field decay heating in XINSs suggests that similar
processes may be occuring in radio pulsars that have ages and magnetic
fields comparable to those of the INSs. Thanks to radio surveys over the
past decade \citep[e.g.][]{mlc+01,mhl+02}, a class of radio pulsar having
inferred magnetic fields comparable to XINSs and even \emph{bona fide}
magnetars has emerged. What physically differentiates these high-$B$ RPPs
from these other classes is unclear; aside from their radio emission, which
was selected for via their discovery in radio surveys, their X-ray
emission properties may well be similar to those of the XINSs and
magnetars. Indeed, in 2009 the High Time Resolution Universe survey
discovered a radio pulsar whose rotational and X-ray properties identified
it as a magnetar \citep{lbb+10}.

Here we present X-ray observations of three high-$B$ radio pulsars:
PSRs J1734$-$3333, B1845$-$19, and J1001$-$5939. These as a group
are key sources for X-ray study, as they share spin, and hence spin-inferred
properties with the XINSs and magnetars. Thus they might be expected
to share X-ray properties, too.

PSR J1734$-$3333 has period $P=1.17\,\mathrm{s}$ and period derivative
$\dot{P}=2.3\times10^{-12}$ and was discovered in the Parkes Multibeam
Survey \citep{mhl+02}. Its spin parameters imply a spin-down luminosity
of $\dot{E}\equiv4\pi^{2}I\dot{P}/P^{3}=5.6\times10^{34}$\,erg\,s$^{-1}$,
characteristic age of $\tau\equiv P/2\dot{P}=8.1$\,kyr, and an inferred
surface dipolar magnetic field of $B\equiv3.2\times10^{19}\bigl(P\dot{P}\bigr)^{1/2}\mathrm{G}=5.2\times10^{13}$\,G,
which is among the highest of all known radio pulsars and similar
to those of \textit{bona fide} magnetars such as 1E 2259+586 ($B=5.9\times10^{13}\,\mathrm{G}$;
\citealt{kcs99}) and Swift J1822.3$-$1606 ($B=4.7\times10^{13}\,\mathrm{G}$;
\citealt{lsk+11,snl+12}; but see \citealt{rie+12}). It has a radio
dispersion measure (DM) of 578\,pc\,cm$^{-3}$ which, based on the
NE2001 model for Galactic free electron density \citep{cl02}, gives
a best-estimated distance to the pulsar of 6.1\,kpc (although these
distance estimates typically have large uncertainties of $~$25\% or more).
Based on its unusually low braking index $\left(n=0.9\pm0.2\right)$,
\citet{elk+11} suggested that the pulsar's magnetic field is growing,
i.\,e.\ its trajectory on a conventional $P/\dot{P}$ diagram is up and to
the right, towards the region occupied by the magnetars. To search for
anomalous X-ray emission from the pulsar, a short \emph{XMM-Newton}
observation was taken in 2009. The source was detected, and although
there were indications that it was anomalously hot, it was too faint
and the observation too short to determine the spectrum unambiguously
or detect pulsations \citep{oklk10}.

PSR B1845$-$19 is a 4.31-s period radio pulsar discovered in the
Molonglo pulsar survey \citep{mlt+78}. It has $\dot{P}=2.33\times10^{-14}$,
$\dot{E}=1.1\times10^{31}$\,erg\,s$^{-1}$, $\tau=2.9$\,Myr,
and $B=1.0\times10^{13}$\,G. It has a DM of 18\,pc\,cm$^{-3}$,
implying a distance of 0.75 kpc \citep{cl02}. The pulsar's magnetic
field, age, $\dot{E}$ and distance are close to those of the XINSs.

PSR J1001$-$5939 was another pulsar discovered in the Parkes Multibeam
Pulsar Survey \citep{lfl+06}, with period $P=7.73$\,s and period
derivative $\dot{P}=5.99\times10^{-14}$. As with PSR B1845$-$19,
the low $\dot{E}$ ($5.1\times10^{30}$\,erg\,s$^{-1}$), age ($\tau=2.1$\,Myr),
and high magnetic field ($B=2.2\times10^{13}$\,G) of PSR J1001$-$5939
are similar to the properties of the known XINSs, although at a distance
of 2.7\,kpc (as determined from its DM of 113\,pc\,cm$^{-3}$)
it is farther away than any of them.

In this paper, we report on X-ray observations of these three high-$B$
RPPs, done with the aim of searching for magnetar- or XINS-like emission
as might be expected if magnetic-field-decay heating is occuring.
Section~\ref{sec:Obs} describes our X-ray observations, Section~\ref{sec:Res}
our results, and we discuss the implications of our findings in Section~\ref{sec:Disc}.

\section{Observations\label{sec:Obs}}

\subsection{PSR J1734$-$3333}

A deep, 125-ks observation of PSR J1734--3333 was carried out on 2011
March 11-12 with the \emph{XMM-Newton} observatory \citep{jla+01}.
We also reanalyzed the EPIC pn data from the 10-ks 2009 \emph{XMM-Newton}
observation of the source \citep{oklk10}, but not the MOS data, as
they had neither the proper time resolution nor sufficient source
counts to meaningfully contribute to our analysis. In both observations,
the EPIC pn camera \citep{sbd+01} was operating in large-window mode
and the EPIC MOS cameras \citep{taa+01} in full-window mode; the
medium filter was in use for all three cameras. Details of the two
observations, as well as those described in Sections \ref{sub:1845obs}
and \ref{sub:1001obs}, are summarized in Table~\ref{tab:Obs}. The
data from both observations were analyzed with the \emph{XMM} Science
Analysis System (SAS) version 11.0.0%
\footnote{See \url{http://xmm.esac.esa.int/sas/}%
}. To search for times of high background flaring that are known to
sometimes affect \emph{XMM-Newton} data, we extracted light curves
from over the entire field of view of all three cameras. The 2011
observation was heavily affected, with over half the exposure length
contaminated by background flares. The 2009 observation showed no
such problems.

\subsection{PSR B1845$-$19\label{sub:1845obs}}

PSR B1845$-$19 was observed for 38\,ks by \emph{XMM-Newton} on 2011
March 16. The pn camera was operating in large-window mode with the
thin filter, and the MOS cameras were operating in full-window mode
also using the thin filter (see Table~\ref{tab:Obs} for further details).
The data were reduced with \emph{XMM} SAS 11.0.0, and times of high
background flaring were removed. Because the two MOS cameras are similar
instruments, we combined their images for the analysis of these data.

\subsection{PSR J1001$-$5939\label{sub:1001obs}}

On 2011 October 8, the \emph{Chandra X-ray Observatory} observed the
position of PSR J1001$-$5939 (see Table~\ref{tab:Obs}) for 19\,ks,
using the back-illuminated S3 chip of the Advanced CCD Imaging Spectrometer
\citep[ACIS;][]{gbf+03} as aimpoint. The 1/8 subarray mode was chosen
to achieve a time resolution of 0.44\,s, sufficient for timing a
pulsar of period 7.73\,s. For our analysis, we worked with the level
2 event files, which were created from level 1 event lists by including
good time intervals as well as information about issues such as bad
pixels and cosmic rays. The data were analyzed using the tools provided
in CIAO%
\footnote{See \url{http://cxc.harvard.edu/ciao/}%
} 4.3.

\section{Results\label{sec:Res}}

\subsection{PSR J1734$-$3333}

\subsubsection{Imaging}

Using the SAS tool \texttt{edetect\_chain} to perform a blind search
for point sources, we detected the X-ray counterpart of PSR J1734$-$3333
in both the 2009 and 2011 observations, with the reported count rates
in both observations being consistent with each other. Our best-fit
position for the X-ray source is the one reported by \texttt{edetect\_chain}
for the 2011 observation: (J2000) $\mathrm{R.A.}=17^{\mathrm{h}}34^{\mathrm{m}}27\fs03(4)$,
$\mathrm{decl.}=-33\degr33\arcmin22\farcs4(5)$, where the listed
uncertainties are the statistical errors only. It is consistent with
the radio timing position of $\mathrm{R.A.}=17^{\mathrm{h}}34^{\mathrm{m}}26\fs9(2)$,
$\mathrm{decl.}=-33\degr33\arcmin20\arcsec(10)$ \citep{elk+11}.
However, \citet{oklk10} also identified an optical source (NOMAD
Catalog%
\footnote{\url{http://www.usno.navy.mil/USNO/astrometry/optical-IR-prod/nomad}%
} ID 0564--0621454) that lay within the 2009 X-ray error circle
and determined that either it or the radio pulsar could be associated
with the X-ray source, but not both. We find that the updated X-ray
position excludes the optical source, even when including \emph{XMM-Newton}'s
$2\arcsec$ absolute pointing uncertainty%
\footnote{See \url{http://xmm2.esac.esa.int/docs/documents/CAL-TN-0018.pdf}%
}, strengthening the conclusion from \citet{oklk10} that the X-ray
source is the counterpart to PSR J1734$-$3333. We also found that
the radial profile of the X-ray source is consistent with the \emph{XMM}
point-spread function (PSF). Therefore there is no evidence of extended
emission.

\subsubsection{Timing Analysis}

To search for X-ray pulsations from PSR J1734$-$3333, we extracted
counts in the 0.5--3\,keV energy band from a region of $30\arcsec$
radius centered on the source. The MOS cameras, operating with a time
resolution of 2.6\,s, were unsuitable for this analysis, so only
the pn data were used. Within the 2011 source region we found 576
counts, and by extracting counts in the same energy range from source-free
regions on the same CCDs, we determined that $352\pm12$ of those
counts were from the background. Following the same procedure for
the 2009 data we determined that of the 150 total counts in the source
region, $100\pm6$ were from background photons. After barycentering
the source region events with the SAS tool \texttt{barycen}, we folded
them into eight phase bins based on a radio timing ephemeris obtained
using the Jodrell Bank 76-m Lovell telescope \citep{elk+11}. Since
the time span covered by the radio ephemeris included the epochs of
both the 2009 and 2011 observations, we could produce a single summed
light curve from both sets of data.

Fitting the folded curve to a straight line gave a best-fit $\chi^{2}$
of 11.1 for 7 degrees of freedom, or a 13\% chance that the folded
curve could result with no signal in the data. Therefore, no significant
pulsations were detected. We also performed additional searches, attempting
to improve the signal-to-noise ratio by using an energy range of 1--2\,keV
or a source region of $20\arcsec$ radius, as well as searching around the
predicted period, but in all cases there were again no detections of pulsations.

To find an upper limit for the pulsed fraction, we simulated event
lists with the same number of total counts as found in the source
region. The simulated signal had a sinusoidal profile with a random
phase and had a user-specified area pulsed fraction, where the latter
is defined as the ratio of the pulsed part of the profile to the entire
profile. We found that a signal with area pulsed fraction of $\sim$0.23
would be detected with $>$$3\sigma$ significance 68\% of the time.
Therefore, because 62\% of the counts in the source region are from
the background, we estimate that the $1\sigma$ upper limit on the
area pulsed fraction of PSR J1734$-$3333 is 0.6 in the 0.5--3\,keV
energy range.

\subsubsection{Spectral Analysis\label{sub:1734Spect}}

We extracted the spectrum of PSR J1734$-$3333 from the 2011 pn and
MOS data using a region of $30\arcsec$ radius centered on the X-ray
source position. Background regions were chosen from nearby, source-free
areas on the same CCD as the pulsar, and response and ancillary response
files were generated with the SAS tasks \texttt{rmfgen} and \texttt{arfgen}.
The pn and MOS spectra were grouped to have a minimum of 25 counts
per bin. We also extracted a spectrum from the 2009 pn data using
the same source and background regions as described in \citet{oklk10},
grouped to have a minimum of 20 counts per bin.

These four spectra were then jointly fit in XSPEC 12.7.0 using simple
power-law and blackbody models, as well as with \texttt{nsa} \citep{pszm95}
and \texttt{nsmax} \citep{hpc08} neutron-star atmosphere models,
all using \texttt{phabs} to model interstellar absorption. The absorbed
power-law model provided a statistically acceptable fit to the data
($\chi^{2}=34.6$ for 44 degrees of freedom), but as it gave an unphysically
steep photon index of 5.2, we do not consider it any further. Best-fit
parameters for the other three models are shown in Table~\ref{tab:1734Spect}.

\begin{figure}
\includegraphics[width=0.75\textwidth]{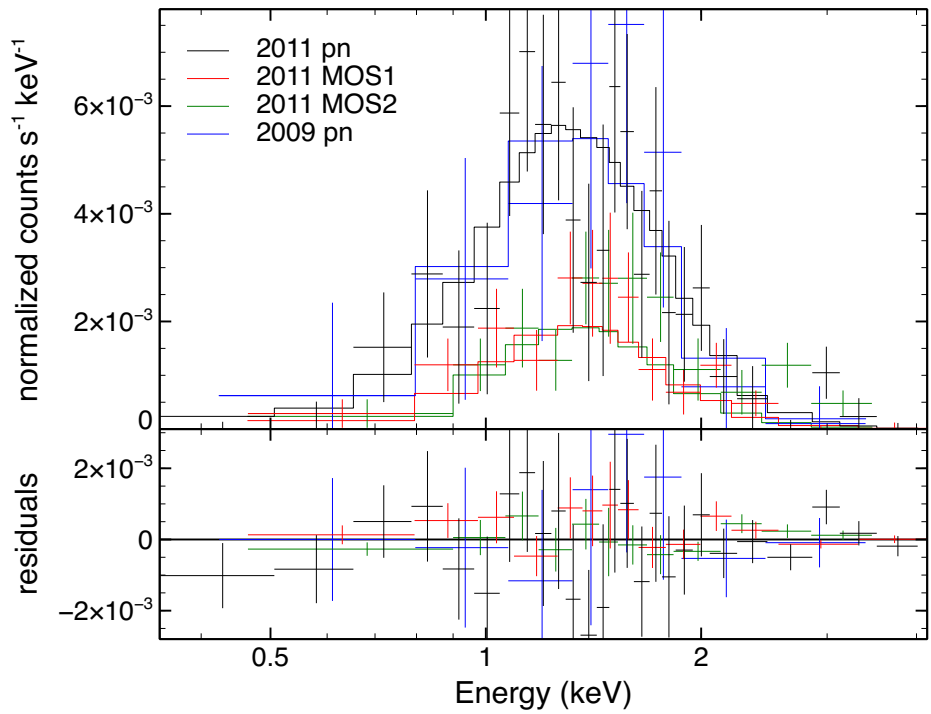}

\caption{\emph{\label{fig:Spect}XMM} spectrum of PSR J1734$-$3333 with the
best-fit blackbody model. The bottom panel shows the residuals in units of
counts\,s$^{-1}$\,keV$^{-1}$. Error bars are at the $1\sigma$ confidence level.}

\end{figure}

\begin{figure}
\includegraphics[width=0.75\textwidth]{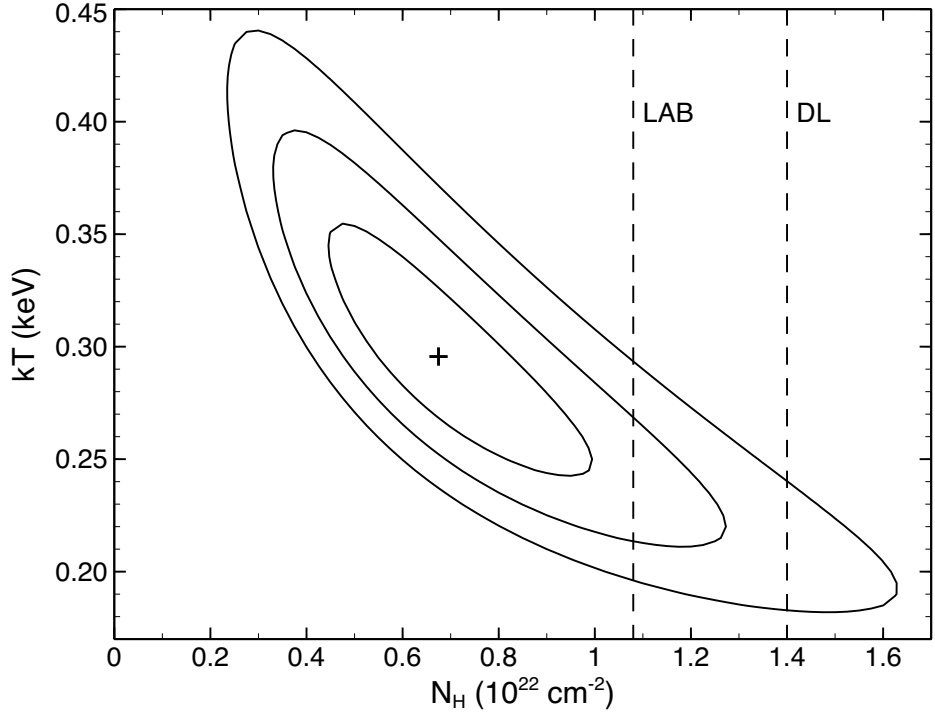}

\caption{\label{fig:Cont}$1\sigma$, $2\sigma$, and $3\sigma$ confidence
contours for $kT$ and $N_{\mathrm{H}}$ from fitting an absorbed
blackbody model to the X-ray spectrum of PSR J1734$-$3333. The cross
marks the best-fit value, and the dashed lines mark the total $N_{\mathrm{H}}$
along the line-of-sight to the pulsar according to the Leiden/Argentine/Bonn
(labeled LAB) and Dickey \& Lockman (labeled DL) Galactic H~\textsc{I}
surveys.}

\end{figure}

The best-fit blackbody model is plotted in Figure~\ref{fig:Spect}.
We found a best-fit%
\footnote{All errors in this section are 90\% confidence intervals.%
} temperature of $kT=0.30\pm0.06$\,keV, corresponding to a column
density of $N_{\mathrm{H}}=0.67_{-0.24}^{+0.36}\times10^{22}$\,cm$^{-2}$.
The blackbody radius is $R_{\mathrm{bb}}=0.45_{-0.20}^{+0.55}\, d_{6.1}$\,km,
giving a bolometric luminosity of $L_{\mathrm{bb}}=2.0_{-0.7}^{+2.2}\times10^{32}d_{6.1}^{2}$\,erg\,s$^{-1}$,
where $d_{6.1}$ is the distance to the pulsar in units of 6.1\,kpc. In
order to better explore the confidence range of $kT$ and $N_{\mathrm{H}}$
for the blackbody model, we plotted their confidence contours in
Figure~\ref{fig:Cont}. The plot shows that the $3\sigma$ lower limit on
$kT$ is 0.18\,keV and that lower values of $kT$ require higher values of
$N_{\mathrm{H}}$. We note that the maximum column density along the line of
sight to the pulsar%
\footnote{\url{http://heasarc.nasa.gov/cgi-bin/Tools/w3nh/w3nh.pl}%
} is $N_{\mathrm{H}}=\left(1.1\textrm{--}1.4\right)\times10^{22}$\,cm$^{-2}$,
and because the DM distance places the pulsar less than halfway through
the Galaxy, the expected column density is significantly lower than
that. As a result, models with $kT<0.2$\,keV which require $N_{\mathrm{H}}>1\times10^{22}$\,cm$^{-2}$
are further disfavored.

Both of the neutron-star atmosphere models provide good fits to the
data and yield much lower best-fit temperatures than the blackbody
model, as is typical. The \texttt{nsa} model gives $kT^{\infty}=0.14\pm0.05$\,keV,
but because it assumes that the emission is from the entire surface of the
star, it implies an unreasonably large distance to the pulsar of $\sim$27\,kpc.
Fixing the distance at 6.1\,kpc results in an even lower temperature
of $90\pm3$\,eV ($\chi^{2}=37.6$ for 45 degrees of freedom), but
the required column density rises to $N_{\mathrm{H}}=1.2_{-0.1}^{+0.2}\times10^{22}$\,cm$^{-2}$,
which is not favored as described above. Given these difficulties,
we also fit the spectrum to the \texttt{nsmax} model for a highly
magnetized neutron star atmosphere. This model gives $kT^{\infty}=0.13_{-0.04}^{+0.05}$\,keV,
$N_{\mathrm{H}}=0.88_{-0.28}^{+0.40}\times10^{22}$\,cm$^{-2}$,
and an emission radius of $R=3.2_{-2.0}^{+8.9}\, d_{6.1}$\,km. Finally, we
note that the \texttt{nsa} and \texttt{nsmax} models were fit using the
highest available values for the magnetic field (in XSPEC), $10^{13}$ and
$2\times10^{13}$~G, respectively. Since these values are below the
$5.2\times10^{13}$\,G field of PSR J1734$-$3333, the results are not
completely reliable.

\subsection{PSR B1845$-$19}

We employed the SAS tool \texttt{edetect\_chain} to search for the
X-ray counterpart of PSR B1845$-$19, but detected no X-ray sources
within $1.5\arcmin$ of the pulsar's radio timing position. From the pn
image, we found 561 counts in a circular region of radius $28\arcsec$
centered on the pulsar position and 781 counts from an annular region of
outer radius $45\arcsec$ around the source region. From the combined MOS
image, we found 132 counts in a $27\arcsec$ radius source region and 192
counts in a surrounding annulus with outer radius $45\arcsec$. In both
cases the source regions were chosen such that they encircle $\sim$80\% of
the energy from the pulsar according to the on-axis PSFs of the
instruments. The background regions are expected to contain a further
$\sim$10\% of the source energy. To estimate the count rate upper limit of
the pulsar, we input these values into the \texttt{aprates} tool from CIAO
4.4 and obtained a $3\sigma$ (99.7\% confidence) upper limit of 0.0048
counts\,s$^{-1}$ for the pn camera and 0.0029 counts\,s$^{-1}$ for the
combined MOS camera. All counts and count rates are for the 0.2--3\,keV
energy band.

In order to convert the count rate upper limit into an upper limit
on the pulsar's surface temperature $kT$, we assume the pulsar's spectrum
can be described by an absorbed blackbody model. We use XSPEC to find the
spectrum for different values of the model parameters (blackbody radius
$R_{\mathrm{bb}}$, distance, and $N_{\mathrm{H}}$) and, based on response
and ancillary response files generated for the source regions described
previously, calculate the predicted count rates for the pn and MOS. We
select 5, 10, and 15~km as our possible values for $R_{\mathrm{bb}}$, and
we use the DM distance to the pulsar of 0.75\,kpc. The total column density
along the pulsar's line of sight\footnotemark[\value{footnote}] is $N_{\mathrm{H}}=\left(0.14\textrm{--}0.18\right)\times10^{22}$\,cm$^{-2}$,
so we evaluate $kT$ upper limits for a range of column densities
with $0\,\mathrm{cm}^{-2}<N_{\mathrm{H}}<0.2\times10^{22}\,\mathrm{cm}^{-2}$
and calculate bolometric luminosity based on $kT$ and $R_{\mathrm{bb}}$.

\begin{figure}
\includegraphics[width=0.49\textwidth]{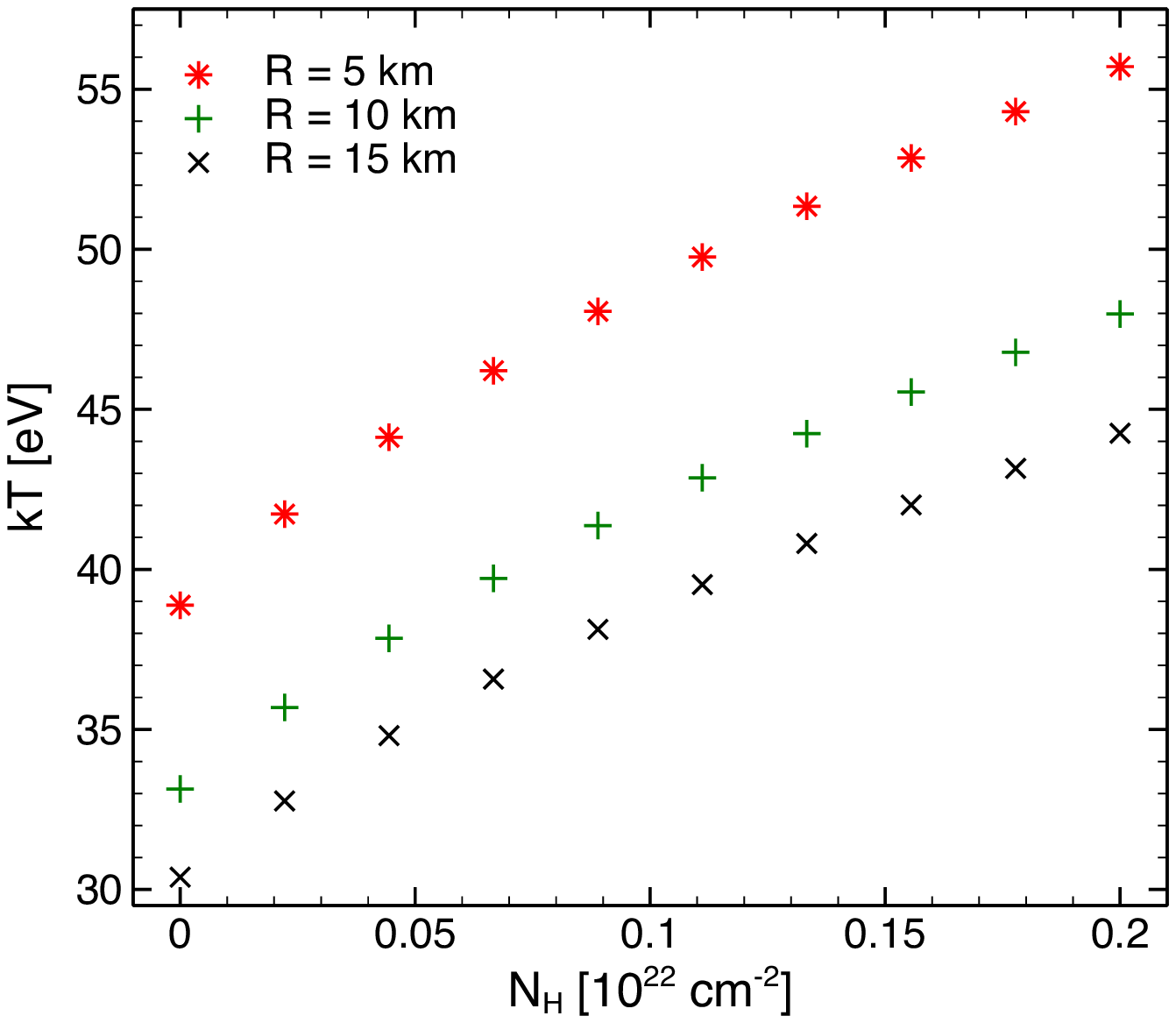}\hfill{}\includegraphics[width=0.42\textwidth]{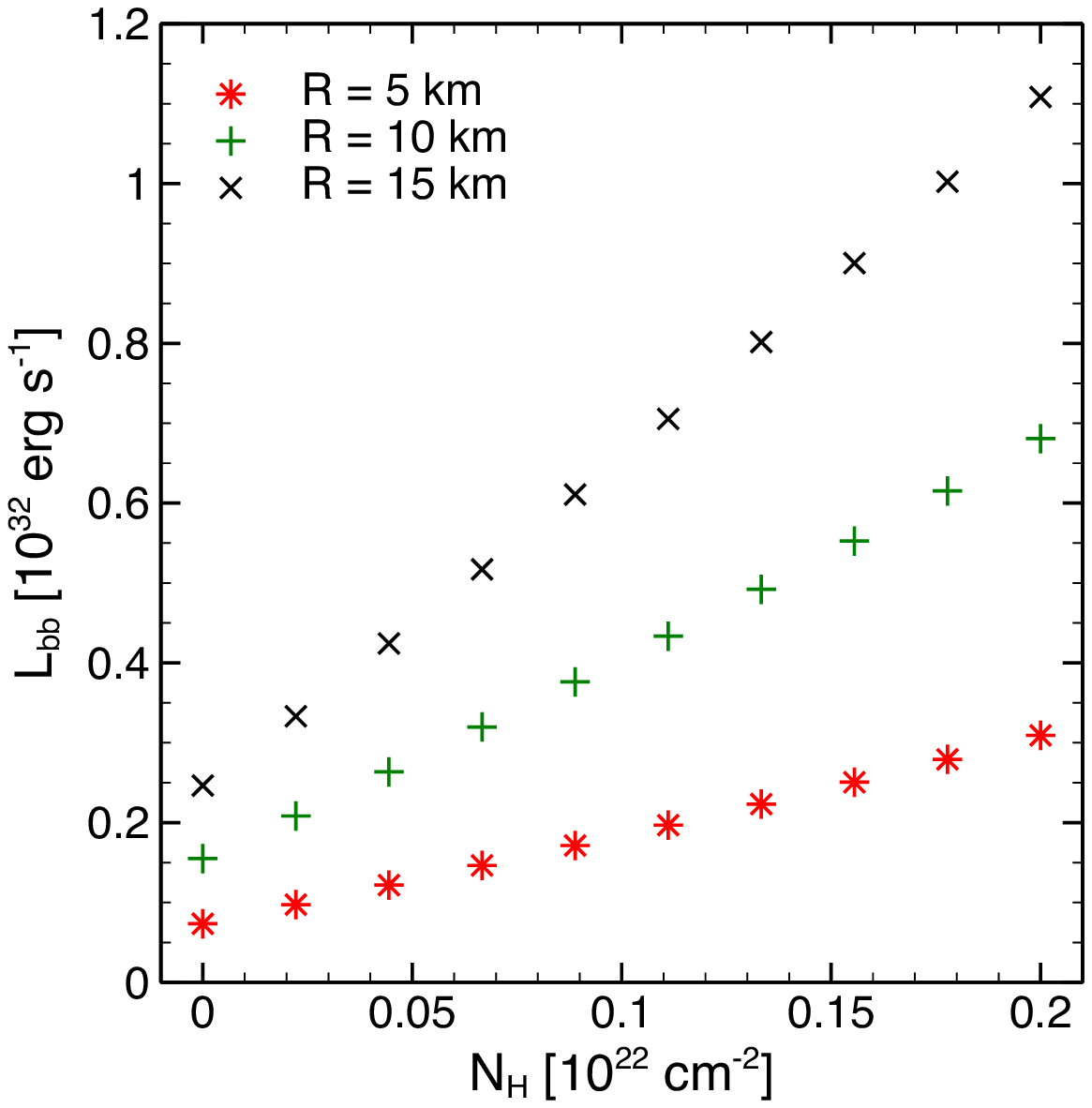}

\caption{\label{fig:B1845}$3\sigma$ upper limits on the blackbody temperature
$kT$ (left) and bolometric luminosity (right) for PSR B1845$-$19.}
\end{figure}

We found that the upper limits from the combined MOS cameras provide
less strict constraints than those of the pn camera, so we present
only the upper limits from the pn image of PSR B1845$-$19 here. These
upper limits, in both $kT$ and bolometric luminosity, are shown in
Figure~\ref{fig:B1845}. Taking the rather conservative values of
$N_{\mathrm{H}}=0.2\times10^{22}$\,cm$^{-2}$ and $R_{\mathrm{bb}}=10$\,km,
our $3\sigma$ upper limits on $kT$ and bolometric luminosity are
48\,eV and $6.8\times10^{31}\mathrm{\, erg\, s^{-1}}\approx6.2\dot{E}$.

\subsection{PSR J1001$-$5939}

No X-ray signal was observed at the expected radio position of PSR
J1001$-$5939. We extracted 0 counts over the 0.3--3\,keV energy
range from a $5\arcsec$ radius circular region centered on the pulsar
position and 141 counts in the same energy range from a surrounding
annular region with outer radius $60\arcsec$. The radius of the source
region was chosen such that it contains $\sim$99\% of the expected
signal based on the \emph{Chandra} on-axis PSF; the rest of the potential
signal should be present in the background annulus. From these values
the CIAO \texttt{aprates} tool gives a $3\sigma$ upper limit on the
pulsar's 0.3--3\,keV count rate of $0.34\times10^{-3}$ counts\,s$^{-1}$.

Using the same procedure as outlined above for PSR B1845$-$19, we
can obtain upper limits for the surface temperature of PSR J1001$-$5939
from our count rate upper limits. We again choose blackbody radii
of $R_{\mathrm{bb}}=5$, 10, and 15\,km, and use the DM distance
to the pulsar of 2.7\,kpc. Based on the DM distance, we also estimate
the column density to be $N_{\mathrm{H}}=0.33\times10^{22}$\,cm$^{-2}$
from the 3-dimensional Galactic extinction ($A_{V}$) model of \citet{dcl03}
and the relation $N_{\mathrm{H}}/A_{V}=1.79\times10^{21}$\,cm$^{-2}$
in \citet{ps95}. Therefore, we evaluate $kT$ and luminosity upper
limits for $0.1\times10^{22}\,\mathrm{cm}^{-2}<N_{\mathrm{H}}<0.55\times10^{22}\,\mathrm{cm}^{-2}$.
Note that the total $N_{\mathrm{H}}$ along the line of sight to the
pulsar\footnotemark[\value{footnote}] is $\left(0.80\textrm{--}0.85\right)\times10^{22}$\,cm$^{-2}$.

\begin{figure}
\includegraphics[width=0.49\textwidth]{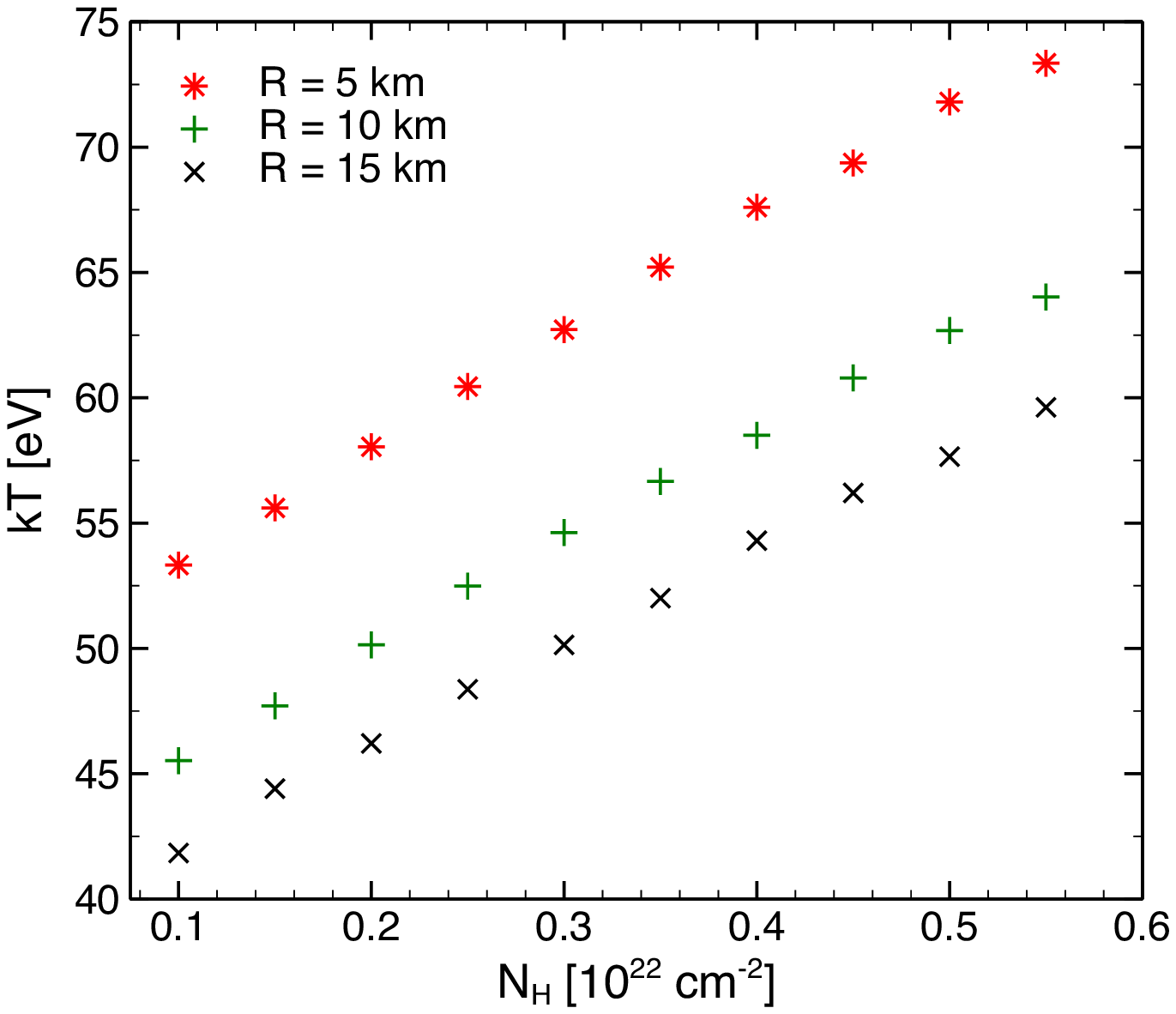}\hfill{}\includegraphics[width=0.42\textwidth]{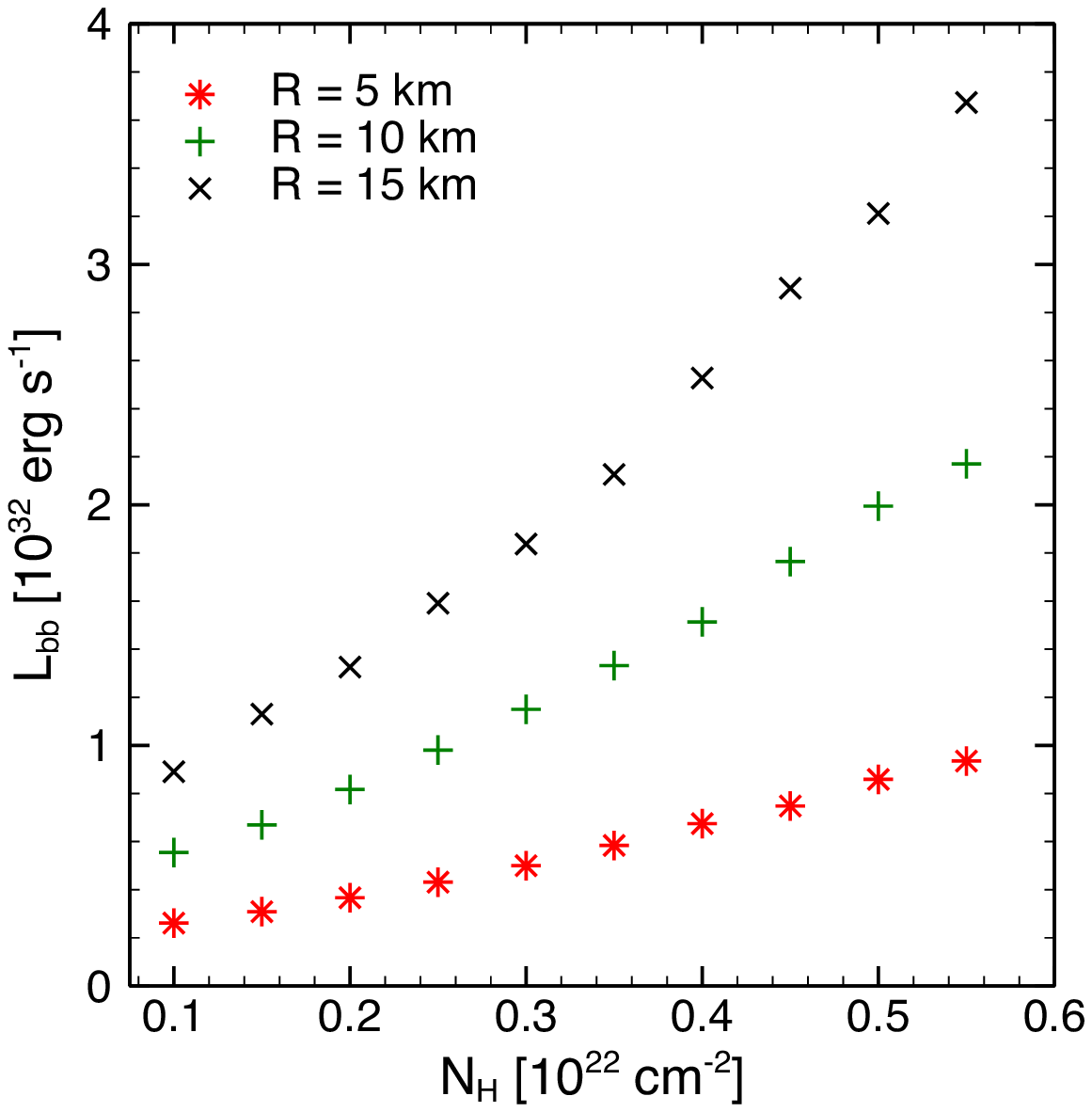}

\caption{\label{fig:J1001}$3\sigma$ upper limits on the blackbody temperature
$kT$ (left) and bolometric luminosity (right) for PSR J1001$-$5939.}
\end{figure}

Figure~\ref{fig:J1001} shows our results for the $kT$ and bolometric
luminosity upper limits of PSR J1001$-$5939. In particular, for our
estimate of $N_{\mathrm{H}}=0.33\times10^{22}$\,cm$^{-2}$ and assuming
a blackbody radius of 10\,km, we find $3\sigma$ upper limits of
56\,eV for $kT$ and $1.26\times10^{32}\mathrm{\, erg\, s^{-1}}\approx25\dot{E}$
for the bolometric luminosity.

\section{Discussion\label{sec:Disc}}

\subsection{PSR J1734$-$3333}

We have reported on a new long \emph{XMM-Newton} observation of the
young, high-$B$ pulsar PSR J1734$-$3333. We found a more precise
position for the X-ray counterpart that falls within the error ellipse
of the pulsar's radio timing position. We were unable to detect X-ray
pulsations, and the $1\sigma$ upper limit of 60\% (0.5--3\,keV)
that we set on the pulsed fraction is not very constraining.

PSR J1734$-$3333 has a thermal spectrum, well fit by a blackbody
model with an unusually high temperature, $kT=0.30\pm0.06$\,keV,
and corresponding blackbody radius of $0.45_{-0.20}^{+0.55}\, d_{6.1}$\,km.
The bolometric luminosity is $L_{\mathrm{bb}}=2.0\times10^{32}d_{6.1}^{2}$\,erg\,s$^{-1}$,
giving an X-ray efficiency of $L_{\mathrm{bb}}/\dot{E}=0.0036d_{6.1}^{2}$.
The pulsar's small emission radius is suggestive of thermal emission
from heated polar caps, but such models predict X-ray efficiencies
less than $10^{-3}$. For example, the polar cap reheating model of
\citet{hm01} predicts $L_{\mathrm{bb}}/\dot{E}\approx3\times10^{-4}$
for PSR J1734$-$3333, an order of magnitude below what we observe.
Of course, if the DM distance is incorrect and the true distance to
the pulsar is within 2\,kpc $\left(d_{6.1}\lesssim0.3\right)$, then
the X-ray efficiency drops to roughly what is predicted for polar
cap reheating. However, \citet{hm01} warn their model may not apply
to pulsars with $B\gtrsim4\times10^{12}$\,G so these predictions
are unreliable.

The pulsar's blackbody temperature of 0.30\,keV is almost three times
as high as expected from a minimal cooling model (0.07--0.11\,keV;
\citealt{pgw06}) given its age. Even the neutron-star atmosphere
models give best-fit temperatures that are too high, $kT^{\infty}=0.13\textrm{--}0.14$\,keV,
though their uncertainties do overlap with the range of predicted
cooling temperatures. For instance, the \texttt{nsa} model with distance
fixed at 6.1\,kpc gives $kT^{\infty}=0.09$\,keV. However, note
that as with the simple blackbody, lower temperatures are correlated
with higher $N_{\mathrm{H}}$ in the atmosphere models too, which
can disfavor that region of parameter space as explained in Section~\ref{sub:1734Spect}.

\begin{figure}
\includegraphics[width=0.9\textwidth]{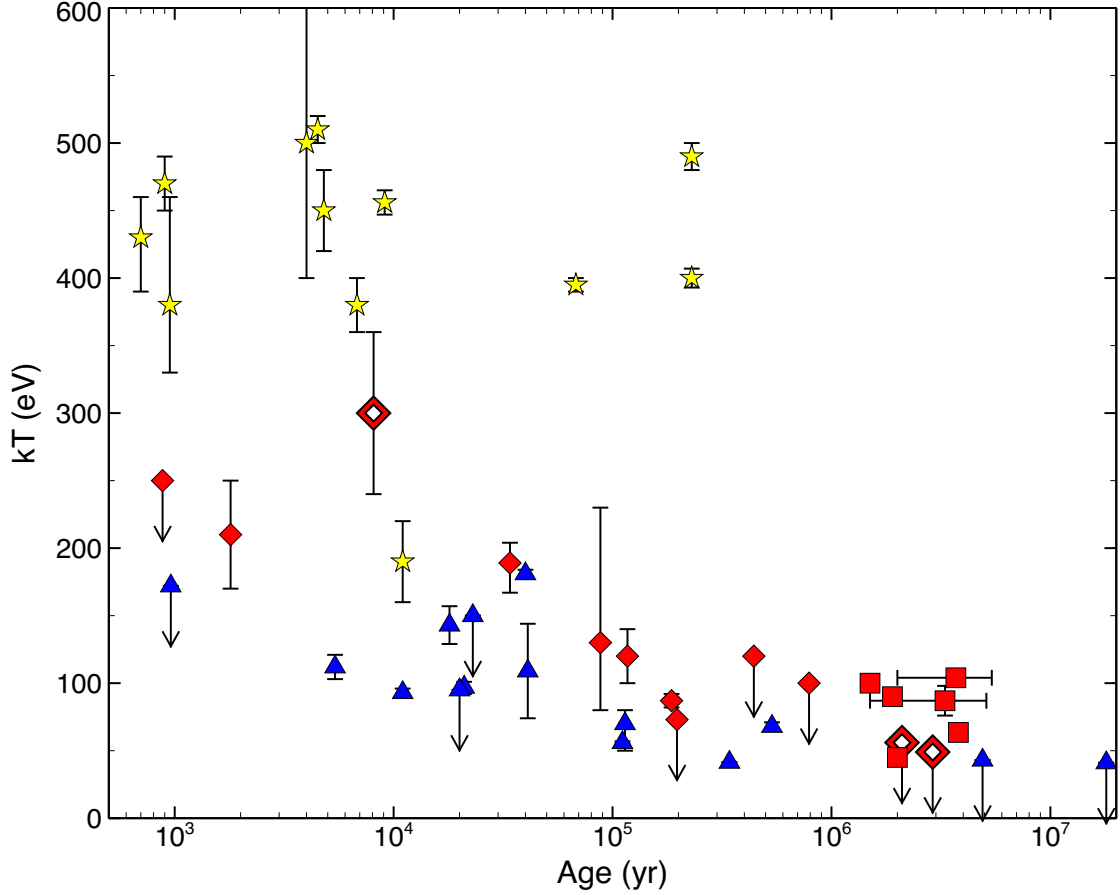}

\caption{\label{fig:AgekT}Blackbody temperatures vs.\ characteristic ages
for normal pulsars (blue triangles), high-$B$ pulsars ($B\geq10^{13}$\,G;
red diamonds), XINSs (red squares), and magnetars (yellow stars).
Magnetars were plotted only if they had a measurement of $kT$ in
quiescence with uncertainties of 0.1\,keV or less (1E~1048.1$-$5937,
1E~1547.0$-$5408, 1E~1841$-$045, 1E~2259+586 1RXS J170849.0$-$400910,
4U~0142+61, CXOU J010043.1$-$721134, CXOU J164710.2$-$455216, CXOU
J171405.7$-$381031, PSR 1622$-$4950, SGR 1900+14, and XTE J1810$-$197),
based on data taken from \protect\url{http://www.physics.mcgill.ca/~pulsar/magnetar/main.html}.
The data for the other sources are taken from \citet{zkm+11} with
the addition of PSR J0726$-$2612 \citep{skv11}, an updated temperature
for PSR J1119$-$6127 \citep{nkh+12}, an updated
timing solution for RX J0420.0$-$5022 \citep{kv11}, and data from
this work. The three high-$B$ pulsars described in this paper are
represented by large red diamonds with holes.}
\end{figure}

Although our ignorance regarding the correct atmosphere model for
PSR J1734$-$3333 precludes meaningful comparisons of its X-ray emission
with models of neutron star cooling, we can nevertheless compare our
results with those of other RPPs of comparable age. We do this in
Figure~\ref{fig:AgekT} where we plot blackbody temperature versus
characteristic age for RPPs, XINSs, and magnetars. High-$B$ sources
are shown in red and yellow (where ``high'' means $B>10^{13}$\,G).
Although a blackbody model neglects atmospheric effects and hence
is almost certainly incorrect for these sources, it provides a simple
spectral parameterization by which to describe all source spectra
and allows a consistent comparison of spectral properties, even if
ultimately not fully physical. Figure~\ref{fig:AgekT} shows that
PSR J1734$-$3333 has a much higher blackbody temperature than nearly
all other radio pulsars for which this quantity has been measured,
with the possible exception of PSR J1119$-$6127, another high-$B$
RPP (\citealt{gkc+05}; Ng et al. 2012, in preparation). It is even
hotter than the magnetar XTE J1810$-$197, although its temperature
is well below those of the other magnetars. In particular, however,
it is far hotter than multiple other RPPs of comparable age but smaller
magnetic field. Indeed, as noted by \citet{zkm+11}, high-$B$ RPPs
in general show a trend toward higher blackbody temperatures when
compared with lower-$B$ RPPs of similar age. This suggests that the
magnetic field affects the observed thermal properties of RPPs, as
is expected if the star is actively heated by field decay. However,
enhanced thermal emission could also be a result of passive effects
such as magnetically altered thermal conductivity \citep{gkp06,pgk07}
without any active field decay. Distinguishing between active and
passive explanations for the enhanced thermal emission in high-$B$
RPPs could be achieved via detailed modelling, or perhaps by observing
other signatures of active field decay, such as magnetar-like bursts
in high-$B$ RPPs. Indeed this has been observed in one source already
\citep{ggg+08}.

\subsection{PSRs B1845$-$19 and J1001$-$5939}

We have reported on X-ray observations of two rotation-powered pulsars,
PSRs B1845$-$19 and J1001$-$5939, taken with the \emph{XMM-Newton}
and \emph{Chandra} telescopes, respectively. In both cases we failed
to detect the pulsars and set upper limits on their $kT$, assuming
a blackbody spectrum. These two sources were selected for observation
because their spin-derived properties are very similar to those of
the XINSs, but as shown in Figure~\ref{fig:AgekT}, this is not the
case in the X-ray band. The $kT$ upper limits we find for our two
pulsars imply that they are significantly cooler than the bulk of
the XINS population, although we cannot exclude the possibility that
their temperatures are similar to that of the coolest known XINS,
RX J0420.0$-$5022 \citep{hmz+04}.

In the magneto-thermal evolution model of \citet{pmg09} and \citet{ppm+10},
the XINSs are born with magnetic fields much higher than what we currently
measure ($\gtrsim$$10^{14}$\,G, possibly including a significant
toroidal component). As they age, the field decays and heats the surface
of the star. A reasonable explanation, then, for PSRs B1845$-$19
and J1001$-$5939 to have much lower temperatures than most of the
XINSs, is that they have similar ages but were born with lower magnetic
fields. They would therefore experience less heating from magnetic-field
decay and their thermal evolution would more closely resemble the
standard cooling curves of non-magnetic neutron stars. This explanation
was also suggested by \citet{kv11} to account for the low temperature
of RX J0420.0$-$5022, and indeed the upper limits on our sources
do allow for the possibility that they and RX J0420.0$-$5022 followed
similar evolutionary tracks.

Another possible explanation is that PSRs B1845$-$19 and J1001$-$5939
do not have similar ages to the XINSs. The above models assume that
the XINSs have true ages of $\sim$0.5--1\,Myr, based on kinematic
age estimates for some of them \citep{tenh11}. These kinematic ages
are smaller than those inferred from spin-down (1.5--4\,Myr). If
the kinematic ages are correct, then it is possible that our radio
pulsar sources have true ages considerably closer to their characteristic
ages (possibly as a result of different evolution as above). In that
case, they have lower temperatures than most or all of the XINSs simply
because they are older and have cooled further. For a related scenario see
\citet{tzp+11}, who, based on magneto-thermal evolution, model the
low-field magnetar SGR 0418+5729 \citep{ret+10} as being $\sim$1--1.5\,Myr
old and having undergone substantial magnetic-field decay.

\section{Conclusion}

In summary, we have presented results from \emph{XMM-Newton} and \emph{Chandra}
observations of three high-$B$ radio pulsars: PSRs J1734$-$3333,
B1845$-$19, and J1001$-$5939. Our observation of PSR J1734$-$3333
was taken to follow up on its detection in 2009. Although we do not
detect X-ray pulsations from the source, we find that it has a thermal
spectrum with $kT=300\pm60$\,eV, significantly higher than predicted
by cooling models. Atmosphere models (\texttt{nsa} and \texttt{nsmax})
yield lower temperatures that are consistent with standard cooling,
but the \texttt{nsa} model in particular predicts too high of a distance
or column density. Comparing PSR J1734$-$3333 with other RPPs, we
find it has a significantly higher blackbody temperature than nearly
every other radio pulsar for which such a measurement has been made,
particularly other RPPs of similar age but lower magnetic field. This
result suggests that a high magnetic field can affect the observed
thermal properties of pulsars, for example by heating of the surface
due to magnetic field decay, or by passive atmospheric effects. PSR
B1845$-$19 and PSR J1001$-$5939 were not detected in our X-ray observations.
Assuming blackbody emission from a 10-km radius, we derive $3\sigma$
upper limits on their temperatures of 48 and 56~eV, consistent with
standard cooling curves. The two radio pulsars have similar spin-derived
properties to the XINSs, but the temperature limits we find are significantly
below the observed temperatures of all the XINSs except RX J0420.0$-$5022.
We conclude that these radio pulsars may have been born with lower
magnetic fields than any of the XINSs (with the possible exception
of RX J0420.0$-$5022), experienced less magnetic-field-decay heating,
and therefore their evolution more closely resembled that of ordinary
radio pulsars.

\acknowledgements{Part of this work was performed under the auspices of the U.S. Department
of Energy by Lawrence Livermore National Laboratory under Contract
No. DE-AC52-07NA27344. VMK holds the Lorne Trottier Chair in Astrophysics
and Cosmology, and a Canada Research Chair, a Killam Research Fellowship,
and acknowledges additional support from an NSERC Discovery Grant,
from FQRNT via le Centre de Recherche Astrophysique du Qu\'ebec and
the Canadian Institute for Advanced Research. Support for this work was
provided by NASA through Chandra award GO1-12083X.}

Facilities: \facility{\emph{CXO} (ACIS)}, \facility{\emph{XMM} (EPIC)}

\begin{deluxetable}{ccccccc}
\tabletypesize{\small}
\tablewidth{0pt}
\tablecolumns{7}
\tablecaption{Summary of X-ray Observations\label{tab:Obs}}
\tablehead{\colhead{Target} & \colhead{Telescope} & \colhead{Obs ID} &
  \colhead{Date} & \colhead{Detector} & \colhead{Time Resolution (s)} &
  \colhead{Exposure (ks)\tablenotemark{a}}}

\startdata
J1734$-$3333 & \textit{XMM} & 0553850101 & 2009 Mar 9 & pn & 0.048 & \phn8.7 \\
& \textit{XMM} & 0653320101 & 2011 Mar 11 & pn & 0.048 & 42.8 \\
&  &  &  & MOS1 & 2.6\phn\phn & 59.8 \\
&  &  &  & MOS1 & 2.6\phn\phn & 64.5 \\
B1845$-$19 & \textit{XMM} & 0653300101 & 2011 Mar 16 & pn & 0.048 & 28.2 \\
&  &  &  & MOS1 & 2.6\phn\phn & 27.3 \\
&  &  &  & MOS2 & 2.6\phn\phn & 27.3 \\
J1001$-$5939 & \textit{Chandra} & 12561 & 2011 Oct 8 & ACIS-S & 0.44\phn & 17.2 \\
\enddata

\tablenotetext{a}{The exposure time is dead-time corrected and has
  intervals of high background flaring removed.}

\end{deluxetable}

\begin{deluxetable}{cccc}
\tablewidth{0pt}
\tablecolumns{4}
\tablecaption{Spectral Models for PSR J1734--3333\label{tab:1734Spect}}
\tablehead{\colhead{Parameter} & \colhead{Blackbody} & \colhead{\texttt{nsa}\tablenotemark{a}} & \colhead{\texttt{nsmax}\tablenotemark{b}}}

\startdata
$N_{H}$ $\left(10^{22}\,\mathrm{cm}^{-2}\right)$ & $0.67_{-0.25}^{+0.35}$ &$0.83_{-0.29}^{+0.41}$ &$0.88_{-0.28}^{+0.40}$ \\
$kT^{\infty}$ (keV) & $0.30\pm0.06$ & $0.14\pm0.05$ & $0.13_{-0.04}^{+0.05}$ \\
$R^{\infty}$ (km) & $0.45_{-0.20}^{+0.55}$ & 13 (fixed) & $3.2_{-2.0}^{+8.9}$ \\
Distance\tablenotemark{c} (kpc) & \nodata & $27_{-20}^{+50}$ & \nodata \\
$\chi^{2}$ (dof) & 34.7(44) & 34.6(44) & 34.3(44) \\
$f_{\mathrm{abs}}$ $\left(10^{-15}\,\mathrm{erg\, s^{-1}\, cm^{-2}}\right)$\tablenotemark{d} & $9.9\pm1.2$ & $9.9_{-1.2}^{+1.1}$ & $9.8_{-1.1}^{+1.2}$ \\
$f_{\mathrm{unabs}}$ $\left(10^{-14}\,\mathrm{erg\, s^{-1}\, cm^{-2}}\right)$\tablenotemark{d} & $3.6_{-1.4}^{+3.7}$ & $5.6_{-2.7}^{+9.9}$ & $6.6_{-3.4}^{+11.6}$ \\
\enddata

\tablenotetext{a}{The \texttt{nsa} model for a pulsar with $B=10^{13}$\,G.}

\tablenotetext{b}{The \texttt{nsmax} model for a pulsar with $B=2\times10^{13}$\,G.}

\tablenotetext{c}{The distance to the pulsar is fit for in the \texttt{nsa} model,
  while in the other two models the DM distance of 6.1\,kpc is used to
  determine the radius of emission.}

\tablenotetext{d}{Absorbed and unabsorbed fluxes are given in the
  0.5--2.0\,keV band.}
\end{deluxetable}

\end{document}